\documentclass[aps,preprint,showpacs]{revtex4}
\usepackage{graphicx}
\usepackage[dvips]{epsfig}

\begin{document}

\draft
\title{Loschmidt echo and stochastic-like quantum dynamics of nano-particles}

\author{V.A. Benderskii}
\affiliation {Institute of Problems of Chemical Physics, RAS \\ 142432 Moscow
Region, Chernogolovka, Russia}

\author{L.A. Falkovsky}
\affiliation{L. D. Landau Institute for Theoretical Physics, RAS,
Moscow, Russia} \affiliation{ Institute of the High Pressure
Physics, RAS, Troitsk, Russia}
\author{E.I. Kats} \affiliation{Laue-Langevin Institute, F-38042,
Grenoble, France}
\affiliation{L. D. Landau Institute for Theoretical Physics, RAS, Moscow, Russia}

\date{\today}

\begin{abstract}

We investigate
time evolution of prepared vibrational state (system) coupled to a reservoir
with dense spectrum of its vibrational states.
We assume that the reservoir has an equidistant spectrum, and the system - reservoir coupling matrix elements are independent
of the reservoir states. The analytical solution manifests three regimes of the evolution for the
system: (I) weakly damped oscillations; (II) multicomponent Loschmidt echo in recurrence cycles;
(III) overlapping recurrence cycles. We find the
characteristic critical values of the system - reservoir coupling constant for the transitions between these regimes.
Stochastic dynamics occurs in the regime (III) due to inevoidably in any real system
coarse graining of time or energy measurements, or initial condition uncertainty.
At any finite accuracy one can always find the cycle number $k_c$
when dynamics of the system for $k > k_c$ can not be determined uniquely from the spectrum, and
in this sense long time system evolution
becomes chaotic.
Even though a specific toy model is investigated here,
when properly interpreted it yields quite reasonable description
for
a variety of physically relevant phenomena, such as complex vibrational dynamics
of nano-particles,
with characteristic inter-level spacing of the order of 10 $cm^{-1}$,
observed by sub-picosecond spectroscopy methods.

\end{abstract}

\maketitle

Experimental studies of vibrational dynamics in various systems
(ranging from relatively small molecules in a gas phase
(near dissociation boundary) or water
molecules clusters confined to interface, through nano-particles
or  large photochromic molecules) reveal in a time domain
$10^{-13} - 10^{-11} \, s$ very rich variety of regimes
\cite{BM98}, \cite{FS00}, \cite{JF02}, \cite{FE03}. Observed in
such systems  seemingly irregular damped oscillation regimes can
not be explained theoretically in the frame work of widely used
models with reservoirs possessing continuous spectra \cite{LC87},
\cite{BM94}, \cite{WE99}. Indeed in the case of a system coupled
to the continuous spectrum reservoir, 
only smooth crossover between coherent
oscillations and exponential decay is possible upon increasing of the coupling.
Moreover generic complex dynamics is observed
in the systems
with characteristic inter-level spacing of the order of $10 \, cm^{-1}$,
when the recurrence cycle period is in the range of sub-picoseconds, where the measurements  \cite{BM98} - \cite{FE03}
are performed.
The dense but discrete spectra with characteristic inter-level spacings of this order
are typical for a wide variety of nano-particles with $10^2 - 10^3$ degrees of freedom.
In such cases one should expect recurrence cycles and much more rich and sophisticated time evolution.
Motivated by these observations, our intent here is to examine
joint system-reservoir evolution, i.e., recurrency cycles, when the energy is flowing back from the reservoir to
the system.

We investigate
a simple (but yet non-trivial) model
of a single level system coupled to a reservoir
with discrete spectrum
\begin{eqnarray}
\label{jl1}
H = \epsilon _s^0 b_s^+ b_s + \sum _{n} \epsilon _n^0 b_n^+ b_n + \sum _nC_n(b_s^+ b_n + b_s b_n^+)
\, ,
\end{eqnarray}
where $\epsilon _s^0$, and $\epsilon _n^0$ are bare (i.e., without interaction)
eigen states of the system and of the reservoir ($b_s^+$, $b_n^+$ are corresponding creation operators),
and $C_n$ are interaction matrix elements.
The Hamiltonian (\ref{jl1}) has non-zero diagonal elements and only one non-zero row and column.
The secular equation to find the eigen values $\epsilon $ of the Hamiltonian reads as
\begin{eqnarray}
\label{jl2}
F(\epsilon ) = \epsilon - \sum \frac{C_n^2}{\epsilon - \epsilon _n^0} = 0
\, ,
\end{eqnarray}
where the energy is measured in units of mean reservoir inter-level spacing, and counted from the unperturbed
system energy $\epsilon _s^0$.

Time dependent amplitudes  satisfy to the corresponding Heisenberg equations of motion
\begin{eqnarray}
\label{jl3}
i \dot {a}_s = \sum _n C_n a_n \, ; \, i \dot {a}_n = C_n a_s + \epsilon _n^0 a_n
\, ,
\end{eqnarray}
supplemented by the initial condition
\begin{eqnarray}
\label{jl4}
a_s(0) = 1\, ; \, a_n(0) = 0
\, .
\end{eqnarray}
The solution of these equations of motion can be represented as
\begin{eqnarray}
\label{jl5}
{a}_s(t) =  \sum _{\epsilon _n}\frac{\exp (i\epsilon t)}{dF/d\epsilon}|_{\epsilon = \epsilon _n}
\,
\end{eqnarray}
where Eq. (\ref{jl5}) is the sum over the residues in the simple
poles $\epsilon = \epsilon _n$ which are the roots of the secular
equation (\ref{jl2}). Because of level repulsion phenomenon, the
dense spectra can be grouped into series of approximately equidistant levels \cite{ME68}.
If the system-reservoir
dynamics is dominated by a single among these series, the
Hamiltonian (\ref{jl1}) can be simplified following an exactly
solvable model proposed long ago by R.Zwanzig \cite{ZW60} of a
system coupled (independent of the reservoir states,
i.e. $C_n \equiv C$) to a reservoir with equidistant spectrum
(i.e., in our notation $\epsilon _n^0 = n$). In fact the Zwanzig
model treats a simplified version of well known Caldeira-Legget
Hamiltonian widely used in condensed matter physics and chemistry
\cite{LC87}, \cite{WE99}. What is lacking, as far as we know, is
an investigation of the recurrence cycle dynamics for the
Caldeira-Legget Hamiltonian. Indeed in the standard  approach the
reservoir spectrum is assumed to be so dense that any recurrence
is irrelevant on a characteristic measurement time scale. However
it is not always the case and many systems (see \cite{BM98} -
\cite{FE03}) lie in the intermediate range of the parameters, with
discrete but dense spectrum of final states. For the ease of
notation let us assume also that the level $n=0$ of the reservoir
is in the resonance with the system energy level $\epsilon _s$.
Thus we get from Eqs. (\ref{jl2}) and (\ref{jl5})
\begin{equation}
\label{jl6}
\epsilon = \pi C^2 \cot (\pi \epsilon )
\, ,
\end{equation}
and
\begin{equation}
\label{jl7} a_s(t) = 2 \sum _{n=0}^{\infty }\frac{\cos \epsilon
_nt}{1 + \pi ^2 C^2 + (\epsilon _n/C)^2} \, .
\end{equation}

Utilizing the Poisson summation formula one can replace the series
terms in Eq. (\ref{jl7}) by their Laplace transforms, as

\begin{eqnarray}\label{jl9}
a_s(t)=\sum_{k=-\infty}^{\infty}\int_
{-\infty}^{\infty}\frac{dn}{1+\pi^2C^2+(\epsilon_n/C)^2} \exp
(-2\pi i kn+i\epsilon_nt)\\ \nonumber
=\sum_{k=-\infty}^{\infty}\int_
{-\infty}^{\infty}\frac{d\epsilon}{\pi^2C^2+(\epsilon/C)^2} \exp
[2\pi i k(\epsilon-\phi)+i\epsilon t] \, ,
\end{eqnarray}
where we replace the integration variable $n\to\epsilon_n $ and
introduce the phase $\phi=\epsilon_n-n$ which can be taken in the
interval (0, 1/2). According to the eigen-value equation (\ref{jl6}), we have
\begin{eqnarray}
\label{jl10} \frac{d\epsilon _n}{dn}=\frac{\pi^2C^2+(\epsilon
_n/C)^2}{1+\pi^2C^2+(\epsilon _n/C)^2} \, ,
\end{eqnarray}
and
\begin{eqnarray}
\label{jl11} \exp (-  2 \pi i \phi ) = \frac{\epsilon _n - i \pi
C^2}{\epsilon _n + i \pi C^2} \, .
\end{eqnarray}
Since $t \geq 0$ the sum in the (\ref{jl9}) includes only a finite
number of the cycle amplitudes $0 \leq k \leq [t/(2\pi )]$ (where
$[x]$ stands for the integer part of $x$). The integrand has two
poles $\epsilon = \pm i \pi C^2$, and only the positive pole
contributes into the integral. Furthermore, the $ k < 0 $ cycles
do not contribute into $a_s(t)$, and also for $k > [t/2\pi ]$ the
pole $i \pi C^2$ lies outside the integration contour in the upper half plane. Combining
everything we end up with
\begin{eqnarray}
\label{jl12}
a_s(t) = \sum _{k=0}^{[t/2\pi ] } a_s^{(k)}(t)
\, ,
\end{eqnarray}
where
\begin{eqnarray}
\label{jl13} a_s^{(k)}(t)= C^2\int _{-\infty }^{+\infty }d\epsilon
 \frac{(\epsilon + i
\pi C^2)^{k-1}} {(\epsilon - i\pi C^2)^{k+1}} e^{ i\epsilon (t - 2
k \pi) }\, .
\end{eqnarray}
The integral  (\ref{jl13}) can be expressed in terms of the
generalized Laguerre polynomials
$L_k^1$ (where $k \geq 1$) \cite{BE53}
\begin{equation}
\label{jl131}
a_s^{(k)}(t)= 2\frac{\tau
_k}{k}L_{k-1}^1(2\tau _k)\exp \left (-\tau _k\right ) \theta(\tau
_k) \, ,
\end{equation}
where 
\begin{equation}
\label{jl132}
\tau _k =  \pi C^2(t - 2 \pi k)
\,
\end{equation}
is the local time for the
$k$-th cycle. The step function $\theta(x)$ enters Eq.
(\ref{jl131}) because $a_s^{(k)}(t)=0$, for $t < 2 k \pi $. Note
that the Poisson summation formula replaces the discrete series of
the poles along the real axis by a single pole on the imaginary
axis. This pole determines decay probability of the initial
quasistationary state \cite{ZE61}. The expressions (\ref{jl12}) -
(\ref{jl131}) describe the time evolution of the system. The
equation (\ref{jl12}) represents dynamics in terms of a certain
superposition of the coherent eigen-frequency oscillations,
whereas the equation (\ref{jl131}) - in terms of the transition
probabilities.

In the limit of weak coupling, $\pi C^2 \ll 1$, the system time evolution is dominated by the coherent
transitions between the resonance states (i.e., $n=0$ in our case).
There are many recurrence cycles within period $2\pi /C$ of these oscillations.
For the strong system - reservoir coupling, $\pi C^2 \gg 1$ the transitions to many reservoir states,
$|n| < \pi C^2$, contributes to the system time evolution. Interference between the transitions suppresses
probability back-flow (from the reservoir to the system).
In the initial cycle $k=0$ we get exponential decay
\begin{equation}
\label{jl14}
a_s^{(0)}(t)=e^{-\Gamma t}
\, ,
\end{equation}
where as one could expect, the exponential decay rate is
determined by the Fermi golden rule
\begin{equation}
\label{jl15}
\Gamma = \pi C^2
\, .
\end{equation}
However (and it is one of our new observations in this paper) in
the following recurrence cycles we find a sort of Loschmidt echo.
This phenomenom occurs due to quantum mechanical synchronization
of the reservoir - system transitions. The Loschmidt echo, we
found, occurs because in the strong coupling limit, back and forth
transitions of many reservoir states which determine system
dynamics, take place not at the same time. We illustrate these
fine structure features of the Loschmidt echo on the Fig. 1, where
$a_s(t)$ is calculated from Eq. (\ref{jl131}) for $C^2=0.3$ and
for $C^2 = 1$. From the known properties of the Laguerre polynomials
one can show that in the cycle $k$ there are
$k$ components of the Loschmidt echo (and the number of zeroes of
the partial amplitude $a_s^{(k)}(t)$ is also $k$). The integral
intensity of the echo within the cycle $k$ is $2/\Gamma $ and
independent of $k$, whereas the width of the echo (related to
oscillation region for the Laguerre polynomials \cite{BE53}) is
$\simeq 4 k$.

Because probability flow between the system and the reservoir states is determined by $\dot {a}_s = - i C \sum _{n} a_n(t)$,
one can describe the system functional space by the canonical variables $a_s(t)$ and $\dot {a}_s(t)$.
For a few initial cycle numbers, the trajectories are almost periodic in the weak coupling limit,
and fill out densely the available phase space upon increasing of the coupling constant (see Fig. 2).
For the higher cycles the trajectories become more and more complex and tangled, that is considered as a sign
of chaotic behavior in classical dynamics \cite{ZA85}, \cite{TA89}.
The entanglement of the trajectories occurs when the echo components of
the mixing for the overlapping cycles.
To characterize this phenomenom one can introduce the critical cycle number $k_c^{(1)} = \pi ^2 C^2$
when the width of the oscillations for the Laguerre polynomials in the amplitudes $a_s^{(k)}(t)$ is equal to the cycle $k$ period.
For the $k > k_c^{(1)}$ the cycles are overlapped.

In the weak coupling limit upon increasing of the $\Gamma $, the Loschmidt echo intensity
increases, while the coherent component of the amplitude decreases. 
We deduce a rough estimate of 
the critical cycle number $k_c^{(2)})$ from the condition that the total duration of the cycles with the decay rate $\Gamma = \pi C^2$
is of the order of the oscillation period $2\pi /C$. The criterion reads as
\begin{equation}
\label{jl16}
k_c^{(2)} \simeq 2 C^{-3}
\, .
\end{equation}

We conclude that there are three regimes of the time evolution which are shown in the Fig. 3.
In the weak coupling limit (I) the coherent oscillations govern the dynamics, for the strong coupling region (II)
- multicomponent Loschmidt echo takes place, and in the region (III) - the cycle overlap occurs.
The crossover line (i) - (II) corresponds to the condition $\Gamma = 1$ also known as irreversibility criterion
for the continuous reservoir spectrum \cite{BJ68}.
In our case (discrete spectrum reservoir) when there are the recurrence
cycles, this criterion means transition from coherent to incoherent behavior \cite{BK02}.
In the region (I) the dynamics is ergodic, whereas in the region (II) it is not, since many rationally independent
frequencies contribute to the time evolution. Note that the (I) - (II) crossover is similar to the known
in classical dynamics \cite{TA89} ergodic - mixing transitions.

The number of zeroes in the strong cycle overlapping condition (i.e., for $k \gg k_c^{(1)}$)
is $\propto k$, like for the non-overlapping cycles, however, only the small fraction ($\sqrt {k_c^{(1)}/k}$)
of these zeroes belongs to the given cycle $k$. All other zeroes come from the previous cycles.
Therefore the interval between the nearest zeroes in the cycle $k$ is determined by a large number of the
previous cycles. If in the cycle $k$ the neighboring in time zeroes come from the previous
cycles $k^\prime $, $k^{\prime \prime }$, then in the next cycle $k+1$ the interval between these zeroes
increases. Moreover, in the cycle $k+1$ at least one more new zero appears between those.
These phenomena (mixing of zeroes and decreasing of the time interval between the neighboring zeroes)
could lead to chaotic long time behavior at any finite accuracy of time or energy measurements (or by other words at any
finite coarse graining of the system). Indeed in such conditions
one may not restore uniquely the wave function from the time dependent amplitudes $a_s(t)$.
To do it one has to know more and more precisely the widely oscillating function.
At any finite accuracy one can always find the cycle number $k_c$
when dynamics of the system for $k > k_c$ can not be determined uniquely from the eigen state spectrum, and
in this sense long time system evolution
becomes chaotic.
It is worth noting a similarity of this phenomenom with quantum mechanical uncertainty
due to quantum object - measurement device interaction \cite{GR93}.

We do believe that our main results (multicomponent Loschmidt echo, three different regimes of the time
evolution, cycle mixing) are valid at least qualitatively for a generic situation
of any simple system coupled to a discrete spectrum reservoir,
because in fact the results do not depend essentially on the details of the spectrum, but only
on its average characteristics (the similar phenomenom is known for nuclear reactions \cite{ME68}).
Note to the same point that modern femtosecond spectroscopy methods (see e.g.,
\cite{BM98} -  \cite{FE03}) indeed demonstrate (in a qualitative
agreement with our consideration) remarkably different types of
behaviors (exponential decay and complicated oscillation) of
relatively close initially excited states.

\acknowledgements

The research described in this publication was made possible in part by RFFR Grants.
Two of us (V.A.B., and L.A.F) have been supported partially by the CRDF grant 2575.
It is our pleasure also to acknowledge helpful discussions with S.P.Novikov.
We thank also V.A.Dubovitskii and L.N.Gak for numerical calculations.


\centerline{Figure captions}

1.Fig. 1

The system time evolution: (a) for $C^2 =0.3$; (b) for $C^2 = 1.0$.

2.Fig. 2

The system trajectories for the overlapping recurrence cycles ($C^2 =1$, $[t/2 \pi ] = 30$).

3.Fig. 3

The phase diagram with three dynamical regimes: (I) - coherent oscillations; (II) - incoherent evolution;
(III) - mixing behavior.

\end{document}